# A Modern Approach to Integrate Database Queries for Searching E-Commerce Product

Ahmad Tasnim Siddiqui (Corresponding author)
College of Computers and Information Technology, Taif University
Taif, Saudi Arabia

Mohd. Muntjir
College of Computers and Information Technology, Taif University
Taif, Saudi Arabia

*Abstract*- **E-commerce refers to the utilization of electronic data transmission for enhancing business processes and implementing business strategies. Explicit components of e-commerce include providing after-sales services, promoting services/products to services, processing payment, engaging in transaction processes, identifying customer's needs, processing payment and creating services/products. In recent times, the use of e-commerce has become too common among the people. However, the growing demand of e-commerce sites have made essential for the databases to support direct querying of the Web page. This re-search aims to explore and evaluate the integration of database queries and their uses in searching of electronic commerce products. It has been analyzed that e-commerce is one of the most outstanding trends, which have been emerged in the commerce world, for the last decades. Therefore, this study was undertaken to ex-amine the benefits of integrating database queries with e-commerce product searches. The findings of this study suggested that database queries are extremely valuable for e-commerce sites as they make product searches simpler and accurate. In this context, the approach of integrating database queries is found to be the most suitable and satisfactory, as it simplifies the searching of e-commerce products.**

## I. INTRODUCTION

The ability of the e-commerce to enable the users to search conveniently for the products in databases is critical to its success. Even though data base queries are considered as the most effective method to access the product database of the e-commerce sites, no significant amount of researches have enlightened the benefits of integrating database queries with e-commerce product searches (Agrawal et al., 2001). In this paper, we have highlighted how e-commerce searches over structured product entities can be optimized by keyword queries such as "iphone 6". However, one major challenge of using database queries for e-commerce product searches is the language gap between the specifications of the products in the databases and the keyword utilized by the people in the search queries (Vander Meer et al., 2012). Google style search box is the most extensively used web interface where the submitted queries neither attribute unit or names.

The intention of this paper is to draw attention towards database queries as well as their use in e-commerce products searches. According to Li and Karahanna (2012), electronic commerce can be explained as the trade of services and/or products, by using internet. E-commerce database queries can be understand as one of the most important database operations, which are totally based on the relational model. The relational models were established by Codd. The term relation is used here in its accepted mathematical sense. Given sets S1 , S2, . . . , Sn, (not necessarily distinct), R is a relation on these n sets if it is a set of n tuples each of which has its first element from S1, its second element from S2 , and so on. It is important to notice that evolution of query form is one of the most effective and integrated user interfaces, which are being used for querying the databases for various applications (Vander Meer, et.al, 2012). An example of the query based searching can be the user searching for a laptop on e-commerce sites with a processor rated around 2.4GHz and around 4 GB of internal memory, can make a query by inserting laptop 2.4 4gb (Agrawal et al., 2001). In this case, it has been assumed that if the e-commerce databases get a close match on numbers then it is possible to get an accurate match for the attribute names. The integration of database queries provides a unified access to various e-commerce search engines (Chiu et al., 2014). These optimized database queries are of great importance since they allow the users to search and compare the products and services of different brands from various sites. To avoid the bottleneck situation of database server and web server we should use the optimized and scalable queries.

## II. AIMS AND OBJECTIVES

This research aims to explore integration and utilize of database queries in e-commerce product searches. The term e-commerce refers to the selling, buying and commerce performed electronically i.e. by the web based services. With the advancements in internet and emergence of new technologies, e-commerce is turning out to be highly popular among people due to its benefits (Poggi, et al., 2012). The concept of e-commerce is the center of majority of the discussion; however, the concept lacks an inclusive definition which can be accepted widely (Chiu et al., 2014).

Generally, e-commerce has been defined as "the tools which enhances the relationship of an organization with its stakeholders". The feature common in both the definitions is the significance of customer relationships in terms of their maintenance and establishment (Grandon et al., 2011). Likewise, the development of an e-commerce site can have a significant impact on the transaction costs (Xiao & Benbasat, 2011). For example, through these websites the organizations can make transactions at a relatively lower time and effort. However, there are certain short comings of transaction based websites (Grandon et al., 2011).

## III. SIGNIFICANCE OF RESEARCH

According to Das-Sarma, et.al, (2014), due to the technological improvements have taken place, there is significant changes in the lives of people, in terms of daily routine works. In this regard, electronic commerce can be measured as one of the most prominent developments, which have been occurred in retailing industry. In the studies of Endrullis, et.al, (2012) it has been documented; e-commerce has been established very fast, across the world. According to the Li and Karahanna (2012), database queries are found to be one of the integrated and important methods, which assist users in searching required products or services, in extensively huge and sophisticated databases. The technique of database queries provides in instant and quick product output. In this manner, it can be stated that the integration of database queries in e-commerce product searches is one of the ultimate advantages towards integrated and viable trading activities.

## IV. REVIEW OF LITERATURE

*A. E-commerce Concept and Characteristics*

According to Li and Karahanna (2012), e-commerce can be explained as number of activities, including exchanging, selling, and buying of products, services, and information, using computer system networks, primarily internet. It is important to observe that the term "commerce", is usually referred to the transactions, which are conducted among business entities.

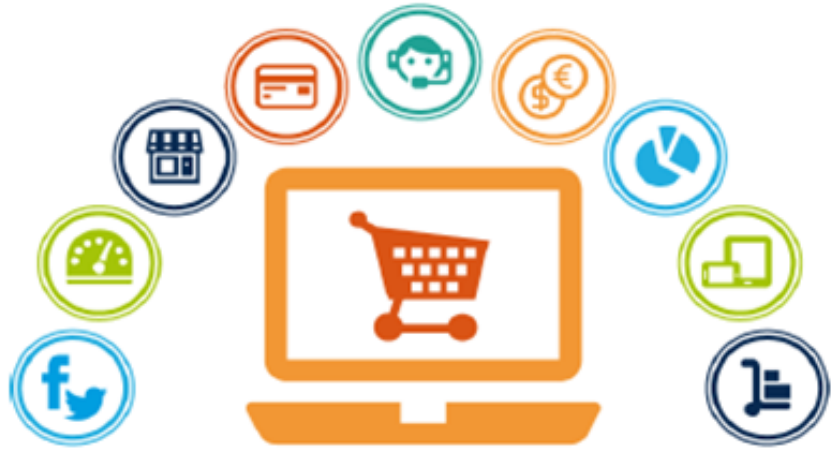

Figure 1. E-commerce (Source: http://www.episerver.com/e-commerce/)

Various practitioners as well as academics have proposed multiple definitions of e-commerce. In simple terms, e-commerce corresponds to online shopping. In terms of business, the definition of e-commerce is not limited to only selling or buying products through internet but also encompasses various processes such collaborating with clients and customers, providing customer service before and after a sale (Lu et al., 2010). The definition of e-commerce proposed by Grandon et al., (2011), summed these processes as a broad array of activities and up and down the value added chain.

Schneider & Perry (2001) defined e-commerce as "the utilization of electronic data transmission in order to enhance business processes and implement business strategies". The term business processes in this definition corresponds to the activities in which the business engage as they attain explicit aspects of commerce. As given by Grandon et al., (2011) explicit components of commerce in relation to supplier include providing after sales services, promoting services/products to services, processing payment, engaging in transaction processes, identifying customer's needs, processing payment and creating services/products. Lu et al., (2010) argued that all these activities or constituents of commerce can be successfully achieved by the means of electronic commerce technologies. However, some of the processes related to business utilize traditional commerce activities in a more effective manner (Lu et al., 2010).

*B. Classification of E-Commerce*

For the purpose of this study, the classification of e-commerce is based upon the business format and business focus. In terms of business focus, the type of business focus is identified by the means of type of buyer, which can be either business clients or end product consumers (Xiao & Benbasat, 2011). In situations, when the buyer is the end consumer, the e-commerce is termed as business-to-consumer e-commerce (B2C). Some websites which are viewed as business-to-consumer e-commerce are eBay.com, Barnesandnoble.com and Amazon.com. On the other hand, when the purchaser is business client or organization, the e-commerce trade is termed as business-to-business e-commerce (B2B) (Fang, 2011).

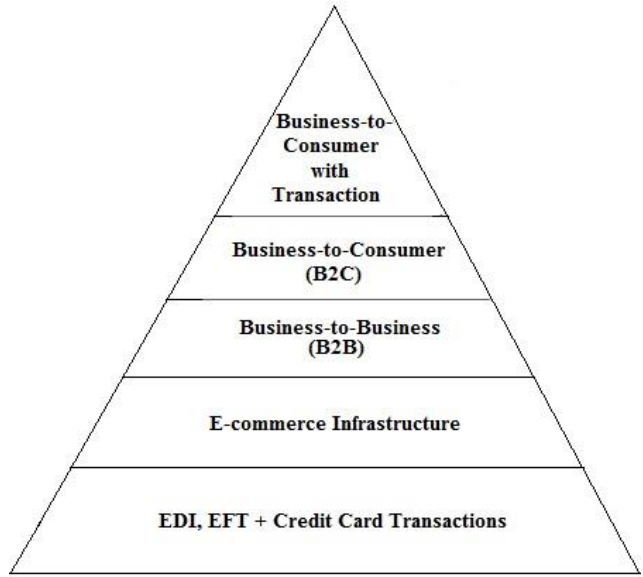

Figure 2. Classification of E-commerce (Source: Fang, 2011)

This model can undertake a range of forms, for instance, there are web based platforms, B2B e-market places and B2B storefronts which gather different retailers and sellers in virtual environment. Some of the chief difference among B2C and B2B are listed below:

- In business-to-business e-commerce, the buyers can purchase larger quantities of desired products.
- In business-to-business e-commerce (B2B), the mode of payment is characterized by purchase order whereas in business to consumer e-commerce (B2C), all the payments are made by credit card regardless of the purchase order.
- In business-to-business e-commerce negotiations are more common and reporting is done by more advanced method (Hinz et al., 2011).
- In business-to-business ecommerce, relationships are considered as extremely crucial.
- In business-to-business e-commerce switching cost is relatively higher.

The figure-2 demonstrates the classification of e-commerce. The virtual teams which are functioning under business-to-business e-commerce are Dell.com and Paper Exchange.com. Although Dell.com also sells its services and products to the consumers, it chief transaction value is achieved through business clients (Hinz et al., 2011). However, it is termed as business-to-consumer e-commerce if it receives its maximum sales from end consumers. In contrast, if the main sale revenue is generated from the business clients, Dell.com is regarded as business-to-business e-commerce.

The e-commerce can also be classified according to its business format. In situations when the chief revenues of e-commerce are generated to online medium, it is termed as an online-dominated channel e-commerce (Casterella & Vijayasarathy, 2013). On the other hand, if the revenues are gained from a non-internet medium, the e-commerce is regarded as a traditional dominated channel e-commerce. Various online sellers such as Amazon.com, eBay.com and Dell.com by selling their products and services online have promoted the concept of e-commerce (Hinz et al., 2011). Subsequently, few auction stores had decided to sell their products through auctions rather than fixed priced deals.

The increasing popularity of online auctions had fostered the creation of various online auction forums such as eBay.com (Geerts & Poggi, 2010). In the recent times, the online auctions have gained so much of popularity that all most all the online stores have initiated a section of auction in which, they sell their products at bargained rates along with fixed price products (Xiao & Benbasat, 2011). All and all, the concept of e-commerce has turned out to be so popular that every service or product desired by the individuals can be found on with negotiable as well as fixed cost.

## V. Database Queries And Integration

Database queries enable the users to retrieve pertinent data from a data. This implies that without searching the entire table, the users can point out the categories of data which can be sought further (Geerts & Poggi, 2010). In addition, database queries also enable the users to merge multiple tables, for instance, if a user is dealing with two tables named as invoices and customers, they can utilize database query to merge the contents of two tables. Subsequently, if a user runs this query, he can attain results which illustrate the name of the customers according to their invoices. It is eminent to mention here that a database query only point out data and does not deals with its storage (Vander Meer et al., 2012). Some of the potential benefits of database query are listed below:

- Merge data from various data sources
- Allows the user to select fields from diversified resources and specify them accordingly
- Identify the records that match the criteria set by the users

*a. Query Reasoning and Expanding Module*

In contrast to traditional query systems, the key attributes of semantic query include expansion of reasoning functions into the queries of user during the querying process (Agrawal et al., 2001). In situations when the semantic information is recommended, the visited records are initially inquired to swiftly identify the interested goods that are allied with the desires of the consumers.

*b. Query Breaking Module*

It is essential to breakdown selected queries further into atomic queries in order to minimize the complexity allied with searching (Vander Meer et al., 2012). This process is mainly carried out by query breaking module. One issue that arises in this module and needs to be resolved is to determine the most appropriate LECO and position at their defined locations.

*c. Integration of Database Queries with E-Commerce*

It is widely recognized that the technology of e-commerce is spreading its horizons rapidly as the buyers are increasingly switching their choices to online stores and markets for purchasing personal care items, clothing and other products or services (Poggi et al., 2012). It is eminent to note here that 6 % of all the revenues are generated through online stores. The midsize sellers are particularly quite interested in e-commerce trading since it allows them to expand their market share and compete with large sized firms. Integrating database queries with e-commerce product searches has numerous benefits (Chiu et al., 2014). For instance, database queries eliminate ambiguity while performing e-commerce product searches. Apart from benefiting e-commerce trading, database queries can be integrated with numerous applications such as custom built apps, personalized marketing apps, e-commerce application, CRM systems and ERP systems.

Database queries when integrated to e-commerce product searches ascertain durability, isolation, consistency and atomicity transactions which are considered crucial for an e-commerce environment (Poggi et al., 2012). Database queries provides e-commerce with the scalability that is required to analyze the data and make decisions concerning the sale channels that need to be invested on the basis of the past performance (Chiu et al., 2014). Customers who make purchases through different e-commerce websites need to able to submit payment information, select product or service they desire to purchase. In parallel, the vendors should be capable of tracking the preferences and inquiries of their customers and process their orders accordingly (Grandon et al., 2011). Therefore, a well organized database query system is required for the development and maintenance of e-commerce website. In case when the web page is static, the content is displayed while the page is being created. However, every time the customer accesses a webpage, the same information is displayed on the static page (Xiao & Benbasat, 2011).

The dynamic web pages which derive little all their content from databases and data files are termed as data based web pages (Lu et al., 2010). These types of web pages are requested when the user press submit button or clicks a hyper link present on the web page form. In certain situations, static query is performed by the programs; for instance, display all items from the inventory. Even though no input is required by the user for this query, the outcomes fluctuate on the basis of the time of query (Fang, 2011). In cases, when the user clicks the submit form button instead of hyper link to make a request, then the form inputs are used by the web server program to create a query. This can be explained with the help of an example in which, the user might chose 10 books to be purchased and subsequently submits input to the web Server programs, which then processes the order and generates a dynamic web Page response to confirm the transaction (Lu et al., 2010).

Cache Early and Cache Often: Implement caching at every layer of your application. Add caching support to the data layer, the business logic layer, and the UI or output layer. Memory is cheap. Caching of data substantially reduces the load on database server. By implementing caching in an intelligent fashion throughout the application, you can achieve great performance gains. Cache can be understood with the diagram given below:

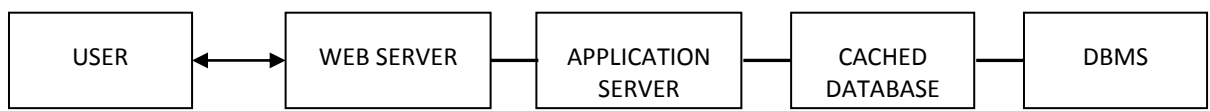

Caching used for an E-commerce Application

Using cache in any e-commerce application directly increases performance of web application.

THE TEST ENVIRONMENT

We created the test application using Visual Studio 2010 and MS-SQL Server 2008. We have used 3-tier application architecture (presentation tier, logic tier and data tier) to test the application. SQL stored procedure was used because of the cached execution plan provides stored procedures a performance advantage over normal inline SQL queries. Stored procedures are more secure. They enhance security. We create more than 67K records in the tables and tried to access the data using web application directly from the server. When we accessed the application after some waiting time we got *System Out Of Memory Exception* show below in figure 4. Same exception was generated when we tried inline SQL queries and also using stored procedure. The query used:

SELECT * FROM PRODUCTS and got the below exception:

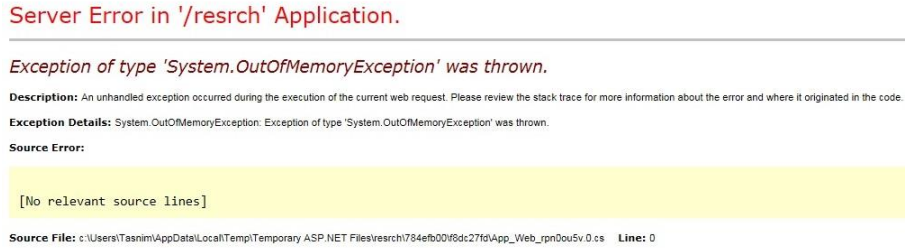

Figure 4. Memory exception error

The sample stored procedure which worked fine with the application is:

CREATE PROCEDURE [dbo].[usp_ProductsListByName]

(@name varchar(20))

AS

BEGIN

SELECT * FROM PRODUCTS WHERE ProdName LIKE '%' + @name + '%'

END

Before running the web application we checked the query retrieval time at T-SQL and we got all the 67216 rows in just 1 second as shown in below figure 5:

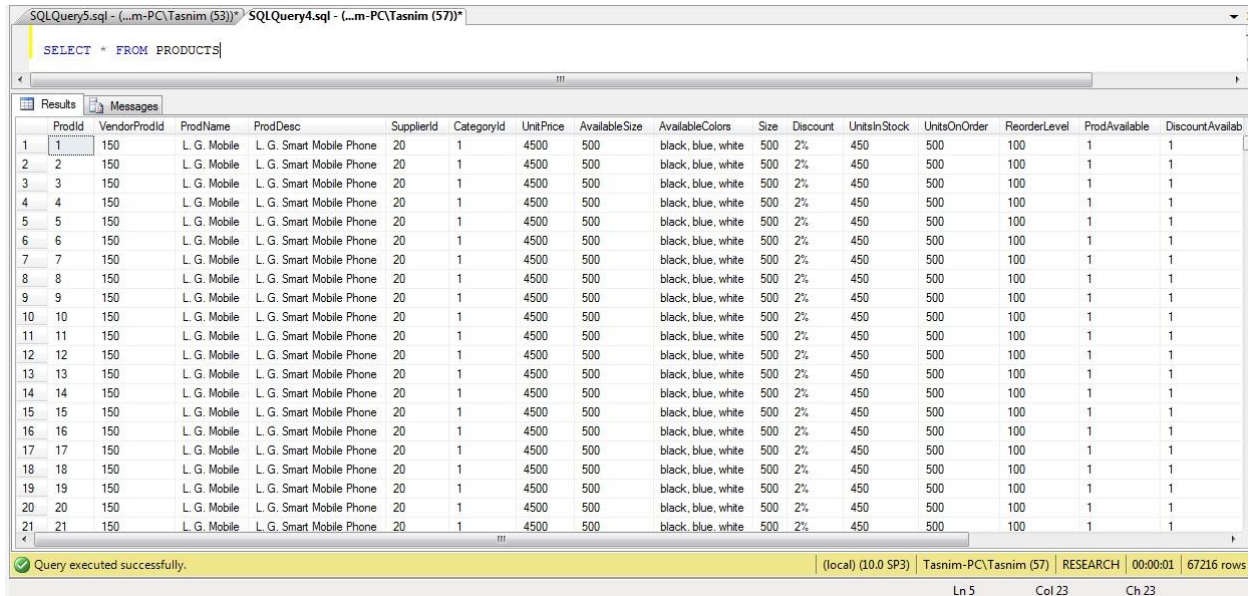

Figure 5. T-SQL query completion time

After optimizing the query used in stored procedure we tried the application for different brands and products and we got the application running successful. For iPhone brand we got the search result of 16000 rows just less than 0.3 seconds as shown in the figure 6 below. The timing was only 0.28301 m. s.

| | | | | | | | | | | | | | |
|---|---|---|---|---|---|---|---|---|---|---|---|---|---|
| ProdId | ProdName | ProdDesc | UnitPrice | AvailableColors | Size | UnitsInStock | UnitsOnOrder | ReorderLevel | ProdAvailable | DiscountAvailable | CurrentOrder | Ranking | Note |
| 402 | iPhone 6 - 64GB | Apple iPhone 6 - 64GB | 55200 | black, white | 500 | 350 | 400 | 100 | True | True | 50 | ***** | Good economical phone |
| 403 | iPhone 6 - 64GB | Apple iPhone 6 - 64GB | 55200 | black, white | 500 | 350 | 400 | 100 | True | True | 50 | ***** | Good economical phone |
| 404 | iPhone 6 - 64GB | Apple iPhone 6 - 64GB | 55200 | black, white | 500 | 350 | 400 | 100 | True | True | 50 | ***** | Good economical phone |
| 405 | iPhone 6 - 64GB | Apple iPhone 6 - 64GB | 55200 | black, white | 500 | 350 | 400 | 100 | True | True | 50 | ***** | Good economical phone |
| 406 | iPhone 6 - 64GB | Apple iPhone 6 - 64GB | 55200 | black, white | 500 | 350 | 400 | 100 | True | True | 50 | ***** | Good economical phone |
| 407 | iPhone 6 - 64GB | Apple iPhone 6 - 64GB | 55200 | black, white | 500 | 350 | 400 | 100 | True | True | 50 | ***** | Good economical phone |
| 408 | iPhone 6 - 64GB | Apple iPhone 6 - 64GB | 55200 | black, white | 500 | 350 | 400 | 100 | True | True | 50 | ***** | Good economical phone |
| 409 | iPhone 6 - 64GB | Apple iPhone 6 - 64GB | 55200 | black, white | 500 | 350 | 400 | 100 | True | True | 50 | ***** | Good economical phone |

Total PRODUCTS: 16000    Total Time Elapsed: 0.283016099998349

Figure 6. SQL Query output for specific product

## VI. CONCLUSION

Coming from all discussions, we can easily conclude that integration of database queries in e-commerce product searches is one of the biggest initiatives towards integrated, constant, and organized e-commerce activities. Nowadays, majority of the people worldwide prefer to use e-commerce sites for purchasing and selling products. These sites are gaining popularity among the users, as they enable to search for the desired products with ease (Geerts & Poggi, 2010). At present, the integration of optimized database queries with the e-commerce product searches is performed either manually or semi automatically (Hinz et al., 2011). The researches being performed with the proliferation of e-commerce, only describe the function and attributes of e-commerce. However, little is known about the benefits of integrating database queries with the e-commerce product searches. Nevertheless, the growing demand of e-commerce raises the need to integrate database queries in its products searches (Hinz et al., 2011). This method makes the search simpler and accurate. The principal purpose of this study was to examine how database queries can be successfully integrated with e-commerce searches. The study also analyzed the core benefits of using e-commerce websites and optimized database queries simultaneously. This research would be valuable for identifying the critical success factors for e-commerce which in turn might assist successful implementation of e-commerce sites in developing countries.

ACKNOWLEDGEMENT

We would like to gratefully and sincerely thank to The Dean of our College, and Chairman of our Departments for their guidance, understanding, patience, and most importantly, their friendly nature during this research paper writing at College of computers and Information Technology, Taif University, Saudi Arabia. We would also like to thank all of the members of the research group and friends who have always supported us for giving the opportunity to write this research paper.